\documentclass[twocolumn,amsmath,amssymb,floatfix,pre,raggedbottom]{revtex4}

\usepackage{graphicx}
\usepackage{dcolumn}
\usepackage{bm}

\begin{document}

\title{Incorporating molecular scale structure into the van der Waals theory of the
liquid-vapor interface}
\author{Kirill Katsov}
\affiliation{Department of Physics, University of Washington\\
P.O. Box 351560, Seattle, WA 98195-1560}
\author{John D. Weeks}
\affiliation{Institute for Physical Science and Technology, and\\
Department of Chemistry and Biochemistry\\
University of Maryland, College Park, MD 20742}
\date{June 10, 2002}
\begin{abstract}
We have developed a new and general theory of nonuniform fluids that naturally
incorporates molecular scale information into the classical van der Waals
theory of slowly varying interfaces. Here the theory is applied to the
liquid-vapor interface of a Lennard-Jones fluid. The method combines a
molecular field treatment of the effects of unbalanced attractive forces with
a locally optimal use of linear response theory to approximate fluid structure
by that of the associated
(hard sphere like) reference fluid. Our approach avoids many of the conceptual
problems that arise in the classical theory and shows why capillary wave
effects are not included in the theory. The general theory and a simplified
version gives results for the interface profile and surface tension for states
with different temperatures and potential energy cutoffs that compare very
favorably with simulation data.
\end{abstract}

\maketitle

\section{ Introduction}
This paper applies our general theory of nonuniform fluids, described in
several earlier publications \cite{1,1.2,2,3,4}, to the liquid-vapor
interface of the simple Lennard-Jones fluid. Our approach here can be viewed
as a generalization of the classical van der Waals (VDW) theory for the density
profile of the liquid-vapor interface \cite{5,6} that (a) incorporates
accurate thermodynamic data for the uniform fluid and (b) corrects the usual
assumption that the interface profile is slowly varying. The new theory
takes account of nonlocal molecular scale density correlations in a very
natural way and can be applied to a wide variety of problems where the
classical theory would fail. This perspective also provides a new and
physically suggestive interpretation of the classical VDW theory for even a
slowly-varying liquid-vapor interface that removes many of the conceptual
problems and ambiguities that arise in standard descriptions \cite{6}. Thus
it seems appropriate to refer to it as a \emph{molecular scale} van der
Waals (MVDW) theory \cite{4}.

Since many aspects of the MVDW theory have been presented in some detail in
previous work, here we will just outline the main features and focus
on the new results we find for the structure and thermodynamics of the
liquid-vapor interface. In this specific application where the interface is
often slowly varying, correction (a) plays the most important role and (b)
is relatively less important, though conceptually (b) represents the most
important advance and permits much more general application of the theory.
In particular we will show how our theory, which first determines the
structure of the liquid-vapor (LV) interface, can be used to calculate
thermodynamic properties such as the surface tension. These results will be
compared to data from computer simulations \cite{7,7a,7b,7c}
and to a simplified version of the
MVDW theory that includes only correction (a). We also discuss the role of
capillary wave fluctuations \cite{6,BLS,15}, which should be taken into account
when comparing theory and experiment.

\section{Molecular field approximation}
An essential ingredient in the MVDW theory and in our interpretation of the
classical VDW theory is the introduction of an effective single particle
potential or ``molecular field'' to describe the locally averaged effects of
the unbalanced attractive forces in the nonuniform LJ fluid \cite{8}. Since
the attractive interactions are relatively slowly varying, such an averaged
treatment seems physically reasonable. To that end, the LJ pair potential $
w(r)\equiv u_{0}(r)+u_{1}(r)$ is separated into rapidly and slowly varying
parts associated with the intermolecular \emph{forces} so that all the
harshly repulsive forces arise from $u_{0}$ and all the attractive forces
from $u_{1}.$ We describe the LV interface using a \emph{grand ensemble}
with a fixed chemical potential $\mu ^{\ell }$ and temperature $k_{B}T\equiv
\beta ^{-1}$ giving two phase coexistence with bulk liquid and vapor
densities $\rho ^{\ell }$ and $\rho ^{v}$ respectively in an external field $%
\phi (\mathbf{r})=0$.

The theory then approximates the structure of the nonuniform LJ system by
that of a simpler nonuniform ``reference'' or ``mimic'' system at the same
temperature but with a chemical potential $\mu _{0}^{\ell }$ and purely
repulsive pair interactions $u_{0}(r)$. These give the same repulsive
intermolecular forces as in the LJ fluid and are well approximated for most
purposes by hard sphere interactions. The nonuniformity in the reference
system is induced by an appropriately chosen \emph{effective reference field}
(ERF) $\phi _{R}(\mathbf{r})$ that is supposed to take into account the
locally averaged effects of the attractive interactions in the LJ fluid.

How should $\phi _{R}(\mathbf{r})$ be chosen? Since we want the reference
fluid structure to approximate that of the full fluid to the extent
possible, it seems reasonable to determine $\phi _{R}(\mathbf{r})$ formally
by the requirement that the local (singlet) densities at every point $%
\mathbf{r}$ in the two fluids are equal \cite{9}: 
\begin{equation}
\rho _{0}(\mathbf{r};[\phi _{R}],\mu _{0}^{\ell })=\rho (\mathbf{r};[\phi
],\mu ^{\ell }).  \label{singletden}
\end{equation}
Of course this density is not known in advance, so in practice we will make
approximate choices for $\phi _{R}$ motivated by mean or molecular field
ideas. Here the subscript $0$ denotes the reference fluid, the absence of a
subscript the LJ fluid, and the notation $[\phi ]$ indicates that the
correlation functions are functionals of the external field $\phi $ (which
in the present case is zero in the LJ system and $\phi _{R}$ in the
reference system). Unless we want to emphasize this point, we will suppress
this functional dependence, e.g., writing Eq.~(\ref{singletden}) as $\rho
_{0}(\mathbf{r)=}\,\rho (\mathbf{r).}$

\section{Classical VDW interface equation}
As discussed in detail in \cite{4}, we can derive the classical VDW
interface equation from this starting point by making two additional
approximations. We briefly discuss this interpretation of the classical
theory and then describe how our new MVDW theory improves on both
approximations.

\subsection{Simple molecular field approximation}
First, the classical VDW theory uses the simple \emph{molecular field }(MF)
approximation for the ERF $\phi _{R}$:
\begin{equation}
\phi _{R}(\mathbf{r}_{1})=\phi (\mathbf{r}_{1})+\int d\mathbf{r}_{2}\,\rho
_{0}(\mathbf{r}_{2};\mathbf{[}\phi _{R}],\mu _{0}^{\ell
})\,u_{1}(r_{12})+2a\rho ^{\ell },  \label{mfeqn}
\end{equation}
where 
\begin{equation}
a\equiv -\frac{1}{2}\int d\mathbf{r}_{2}\,u_{1}(r_{12})  \label{aint}
\end{equation}
corresponds to the attractive interaction parameter $a$ in the uniform fluid
VDW equation, as discussed below. This is just a transcription of the usual
molecular field equation for the Ising model to a continuum fluid with
attractive interactions $u_{1}(r)$ and can be derived in a number of
different ways \cite{10,11}. The connection to the unbalanced attractive forces
is perhaps most clearly seen in the derivation in \cite{1,2}, which starts
from a formally exact description of the force balance in a nonuniform fluid
and arrives at Eq.~(\ref{mfeqn}) by a series of physically motivated
approximations.

For the LV interface we have $\phi (\mathbf{r})=0$, but it is convenient
in what follows to keep a general $\phi $ which
we will then set to zero. In that case we will also choose $\mu _{0}^{\ell }$
so that the density $\rho _{0}^{\ell }$ of the uniform reference fluid with
$\phi _{R}=0$ equals $\rho ^{\ell }$. With this choice the ERF$\phi _{R}$
vanishes on the liquid side far from the interface where the density becomes
equal to $\rho ^{\ell }.$

Another special case of Eq.~(\ref{mfeqn}) arises when $\phi $ is a
\emph{constant}. Since a constant field in the grand ensemble is equivalent to a
shift of the chemical potential, Eq.~(\ref{mfeqn}) then relates the chemical
potentials in the \emph{uniform} LJ and reference fluids \cite{2}.
Equation~(\ref{mfeqn}) thus yields the familiar uniform fluid VDW result \cite{6} 
\begin{equation}
\mu (\rho )=\mu _{0}(\rho )-2\rho a,  \label{mua}
\end{equation}
where $\mu (\rho )$ and $\mu _{0}(\rho )$ denote the chemical potential as a
function of density $\rho $ for the uniform LJ fluid and the reference fluid
respectively. In the classical theory $\mu (\rho )$ is defined for all
$\rho $ in terms of
known reference system quantities by the right side of this equation. Since
the uniform reference fluid is well defined for all densities below freezing,
no problems arise from densities in the two phase region of the LJ fluid. The
MVDW theory will use a slightly different expression for $\phi _{R}$ in Eq.~(%
\ref{mmfint}) below that gives a more accurate description of the
thermodynamics of the uniform LJ\ fluid.

In general, to calculate $\phi _{R}$ a self-consistent solution of Eq.~(\ref
{mfeqn}) is required, since $\phi _{R}$ appears explicitly on the left side
and implicitly on the right side through $\rho _{0}(\mathbf{r};\mathbf{[}
\phi _{R}],\mu _{0}^{\ell })$. Thus a useful implementation of the MF idea
requires a way to accurately determine the density response $\rho _{0}(%
\mathbf{r};\mathbf{[}\phi _{R}],\mu _{0}^{\ell })$ induced by a given $\phi
_{R}.$

\subsection{Local response to ERF}
The classical VDW interface equation results when a \emph{second} approximation,
appropriate only for a slowly varying field, is used to estimate the density
response $\rho _{0}(\mathbf{r};\mathbf{[}\phi _{R}],\mu _{0}^{\ell })$. This
\emph{hydrostatic approximation} for the density takes account only of the 
\emph{local value} of the field through a shift in the chemical potential 
\cite{1.2,3,4}. Thus $\rho _{0}(\mathbf{r}_{1};\mathbf{[}\phi _{R}],\mu
_{0}^{\ell })$ is approximated for each $\mathbf{r}_{1}$ by $\rho _{0}^{%
\mathbf{r}_{1}}$, the \emph{local hydrostatic density response}, which
satisfies
\begin{equation}
\mu _{0}(\rho _{0}^{\mathbf{r}_{1}})=\mu _{0}^{\ell }-\phi _{R}(\mathbf{r}
_{1}).  \label{hydromudef}
\end{equation}
Hence the nonuniform density $\rho _{0}(\mathbf{r}_{1}\mathbf{)}$ at each
$\mathbf{r}_{1}$ is assumed to equal
$\rho _{0}^{{\bf r}_{1}}\equiv \rho _{0}({\bf r}_{1};[0],\mu _{0}^{{\bf r}_{1}})$,
the density of the \emph{uniform} reference fluid in zero field at the
shifted chemical potential $\mu _{0}^{{\bf r}_{1}}\equiv \mu _{0}^{\ell }-%
\phi _{R}(\mathbf{r}_{1})$, given by the right side of Eq.~(\ref{hydromudef}).
When $\phi _{R}(\mathbf{r}_{1})$ is a constant, this gives the exact result.

The superscript
$\mathbf{r}_{1}$ in $\rho _{0}^{\mathbf{r}_{1}}$ is meant to remind us that
$\rho _{0}^{\mathbf{r}_{1}},$ like $\rho _{0}^{\ell }$ or $\rho _{0}^{v}$,
represents the density of the uniform reference fluid at a particular
chemical potential $\mu _{0}^{{\bf r}_{1}}$, which from Eq.~(\ref{hydromudef})
depends parametrically
on $\mathbf{r}_{1}$ through the local value of the ERF. Thus when
$\mathbf{r}_{1}$ is in the bulk liquid (vapor) phase then
$\rho _{0}^{\mathbf{r}_{1}}$ reduces to $\rho _{0}^{\ell }$ ($\rho _{0}^{v}$).
When $\phi _{R}$ is
very slowly varying this ``local field'' approximation is quite accurate and
in this special case is equivalent to the local density approximation made in the usual
interpretation of the VDW theory \cite{6,11}. However this approximation
ignores the nonlocal excluded volume correlations that can be induced by a
more rapidly varying $\phi _{R}$. This represents a major limitation of the
classical theory in more general applications, and will be corrected in the
MVDW theory.

\subsection{Classical interface equation}
\label{responsetoERF}The classical VDW interface equation follows
immediately when $\rho _{0}(\mathbf{r}_{2}\mathbf{)}$ is replaced by $\rho
_{0}^{\mathbf{r}_{2}}$ in Eq.~(\ref{mfeqn}) and the latter is substituted
into Eq. \ref{hydromudef}. This yields an integral equation for $\rho _{0}^{%
\mathbf{r}_{1}},$ which from Eq.~(\ref{singletden}) is supposed to equal the
density in the full LJ fluid: 
\begin{equation}
\mu _{0}(\rho _{0}^{\mathbf{r}_{1}})=\mu _{0}^{\ell }-\phi (\mathbf{r}
_{1})-\int d\mathbf{r}_{2}\,\rho _{0}^{\mathbf{r}_{2}}\,u_{1}(r_{12})-2a\rho
^{\ell }.  \label{vdwref1}
\end{equation}
This can be exactly rewritten in a more standard form using $\mu (\rho )$
as defined in Eq.~(\ref{mua}): 
\begin{equation}
\mu (\rho _{0}^{\mathbf{r}_{1}})=\mu ^{\ell }-\phi (\mathbf{r}_{1})-\int d%
\mathbf{r} _{2}\,[\rho _{0}^{\mathbf{r}_{2}}\,-\rho _{0}^{\mathbf{r}
_{1}}]u_{1}(r_{12}).  \label{vdwref}
\end{equation}
Specializing to the case of the LV interface with planar symmetry and $\phi
=0$ and expanding $\rho _{0}^{\mathbf{r}_{2}}$ to second order in a Taylor
series about $\rho _{0}^{\mathbf{r}_{1}}$ (consistent with the assumption of
a slowly varying interface) yields the classical VDW differential equation
for the interface profile $\rho _{0}^{z}$: 
\begin{equation}
\mu (\rho _{0}^{z})-\mu ^{\ell }=m\,\frac{d^{2}\rho _{0}^{z}}{dz^{2}},
\label{vdwrefode}
\end{equation}
where 
\begin{equation}
m\equiv -\frac{1}{6}\int d\mathbf{r\,}r^{2}u_{1}(r).  \label{vdwm}
\end{equation}

Equations (\ref{vdwref}) and (\ref{vdwrefode}) are completely equivalent to
the VDW theory for the LV interface as it is usually presented \cite{6}. In
our derivation the theory describes \emph{hydrostatic} densities in the
\emph{reference system}, and $\mu (\rho )$ is also defined in terms of
reference system quantities given on the right side of Eq.~(\ref{mua}).
This provides a
simple and consistent interpretation of the classical theory that avoids all
the conceptual problems associated with densities in the two phase region of
the LJ fluid that arise in traditional approaches.

In this derivation we have obtained the VDW interface equation directly,
without first approximating the free energy. We will later show how to
determine the interface free energy in this approach. First we will discuss
our new MVDW theory for the interface profile, which improves on both
approximations made in the classical theory.

\section{MVDW theory for the LV interface profile}
In the limit of a \emph{uniform} system, Eq.~(\ref{mfeqn}) describes all
effects of attractive interactions in terms of the \emph{constant} parameter 
$a$ as in the van der Waals equation. While this very simple approximation
captures much essential physics and gives a qualitative description of the
uniform fluid thermodynamic properties it certainly is not quantitatively
accurate. In particular, when used to describe a slowly varying liquid-vapor
interface, it will predict shifted (molecular field) values for the
densities of the coexisting bulk liquid and vapor phases. The main problem
with the classical theory in this case is not so much its description of the
local density gradients, which are often small, but its predictions for the
thermodynamic properties of the coexisting \emph{bulk} phases. The first
correction made in the MVDW theory is to modify Eq.~(\ref{mfeqn}) so that it
agrees with a given equation of state for the uniform system while still
giving reasonable results for nonuniform systems \cite{2}.

\subsection{Modified molecular field approximation}
To achieve quantitative agreement with known thermodynamic properties of the
uniform LJ\ system we can replace the constant $a$ by a \emph{function} $%
\alpha $ that depends (hopefully weakly, to the extent the van der Waals
theory is reasonably accurate) on temperature and density \cite{2}. Thus,
instead of using the MF \emph{approximation} for $\mu (\rho )$ as in Eq.~(%
\ref{mua}), we assume that $\mu (\rho )$ is known from an accurate bulk
equation of state. In particular, we determine $\mu (\rho )$ from the
33-parameter equation of state for the LJ fluid given by Johnson, et al \cite
{13}. This provides a very good global description of the stable liquid and
vapor phases in the LJ fluid and provides a smooth interpolation in between
by using analytic fitting functions. Thus it naturally produces a modified
``van der Waals loop'' in the two phase region and seems quite appropriate
for our use here in improving the simplest MF description of the uniform
fluid.

Now we relate this accurate $\mu (\rho )$ to the known $\mu _{0}(\rho )$
through a function $\alpha (\rho )$ defined so that 
\begin{equation}
\mu (\rho )=\mu _{0}(\rho )-2\rho \alpha (\rho ).  \label{mualpha}
\end{equation}
Thus the exact chemical potentials in the uniform LJ and reference systems
are related in the same way as is predicted by the simple MF approximation
of Eq.~(\ref{mua}) except that the constant $a$ is replaced by a
(temperature and density dependent) function $\alpha (\rho )$ chosen so that
Eq.~(\ref{mualpha}) holds. We showed in \cite{2} that the ratio $\alpha
(\rho )/a$ is indeed of order unity and rather weakly dependent on density
and temperature.

Because of the strictly local response in Eq.~(\ref{hydromudef}), these
results for a constant field can also be used to determine exact results in
the hydrostatic limit of a very slowly varying field. We want to modify
Eq.~(\ref{mfeqn}) so that in the hydrostatic limit it will reproduce these
exact values, while still giving reasonable MF results for more rapidly
varying fields.

There is no unique way to do this, but the following simple prescription
seems very natural, and gives our final result, which we have called the 
\emph{modified molecular field} (MMF) approximation for the ERF \cite{2}: 
\begin{eqnarray}
\phi _{R}({\bf r}_{1})-\phi ({\bf r}_{1}) &=&\frac{\alpha (\rho _{0}^{{\bf r}%
_{1}})}{a}\int d{\bf r}_{2}\,\rho _{0}({\bf r}_{2};{\bf [}\phi _{R}],\mu
_{0}^{\ell })\,u_{1}(r_{12})  \nonumber \\
&&+2\alpha (\rho ^{\ell })\rho ^{\ell }.  \label{mmfint}
\end{eqnarray}
Thus the molecular field integral in Eq.~(\ref{mfeqn}) is multiplied by a
factor $\alpha (\rho _{0}^{\mathbf{r}_{1}})/a$ of order unity that depends
on $\mathbf{r}_{1}$ through the dependence of the hydrostatic density $\rho
_{0}^{\mathbf{r}_{1}}$ on the local value of the field $\phi _{R}(\mathbf{r}
_{1}),$ and the constant $2a\rho ^{\ell }$ is replaced by the appropriate
limiting value of the modified integral. The MVDW theory assumes that the
ERF is given by Eq.~(\ref{mmfint}) rather than Eq.~(\ref{mfeqn}).

\subsection{Nonlocal response to the ERF}
The second important correction made in the MVDW theory is to determine more
accurately the full \emph{nonlocal} response $\rho _{0}(\mathbf{r}_{1};%
\mathbf{[}\phi _{R}],\mu _{0}^{\ell })$ to the ERF, thus correcting the
local hydrostatic density $\rho _{0}^{\mathbf{r}_{1}}$ used in the classical
theory. We introduced in \cite{1.2} a simple and generally very accurate
method for calculating the structure and thermodynamics of the reference
fluid in the presence of a \emph{general} external field, using linear response
theory in a locally optimal way to calculate the nonlocal corrections to the
hydrostatic density. For a very slowly
varying field the theory gives the hydrostatic density and for a hard core
field the theory naturally reduces to the Percus-Yevick approximation \cite
{14}.

The result is an integral equation for the density $\rho _{0}(\mathbf{r%
}_{1}),$ which we refer to as the hydrostatic linear response (HLR)
equation: 
\begin{equation}
\rho _{0}(\mathbf{r}_{1})=\rho _{0}^{\mathbf{r}_{1}}+\rho _{0}^{\mathbf{r}
_{1}}\int \!d\mathbf{r}_{2\,}c_{0}(r_{12};\rho _{0}^{\mathbf{r}_{1}})[\rho
_{0}(\mathbf{r}_{2})-\rho _{0}^{\mathbf{r}_{1}}].  \label{HLR}
\end{equation}
Here $c_{0}(r_{12};\rho _{0}^{\mathbf{r}_{1}})$ is the direct correlation of
the \emph{uniform} reference fluid at the hydrostatic density $\rho _{0}^{%
\mathbf{r}_{1}}$. This can be accurately approximated using known results for
the uniform hard sphere fluid, as discussed in \cite{1.2,3}.
The $\mathbf{r}_{1}$ dependence of the linear response
kernel $c_{0}$ through $\rho _{0}^{\mathbf{r}_{1}}$ is the most important
new feature of the HLR equation. A discussion of the ideas leading
to Eq.~(\ref{HLR}) and of its advantages
over standard methods, along with numerical details of its solution,
is given in \cite{1.2,3}.

\subsection{Two step method and the MVDW theory}
The MVDW theory for the LV interface arises from the self-consistent
solution of Eqs.~(\ref{hydromudef}), (\ref{mmfint}) and (\ref{HLR}). A
\emph{two-step iterative method} proved sufficient in all cases tested. Given a
starting guess for $\phi _{R}(\mathbf{r}_{1})$ one computes in the first
step the local hydrostatic density response $\rho _{0}^{\mathbf{r}_{1}}$
from Eq.~(\ref{hydromudef}). Then in a second step nonlocal corrections
leading to $\rho _{0}(\mathbf{r}_{1})$ are determined from (\ref{HLR}), and
this is used in Eq.~(\ref{mmfint}) to give a new estimate for $\phi _{R}(%
\mathbf{r}_{1}).$ This process is iterated to self-consistency, and accurate
numerical results are readily obtained.

\subsection{Simplified MVDW interface equation}
A simplified version of the MVDW theory arises when one skips the second
step and assumes that $\rho _{0}(\mathbf{r}_{1})=\rho _{0}^{\mathbf{r}_{1}}$
as in the classical theory, while still using the accurate equation of state
to determine $\mu (\rho )$ from Eq.~(\ref{mualpha}). While the full MVDW
theory is straightforward to implement, the nonlocal corrections
for the LV interface
are often small and the use of the hydrostatic approximation allows for a
more direct comparison with the classical theory. Using this approximation
and Eqs.~(\ref{mmfint}), (\ref{hydromudef}), and (\ref{mualpha}), we obtain
an integral equation analogous to the classical equation (\ref{vdwref}).
Expanding for simplicity to second order and assuming planar symmetry yields
a generalization of the classical interface equation (\ref{vdwrefode}): 
\begin{equation}
\mu (\rho _{0}^{z_{1}})-\mu ^{\ell }=m_{\alpha }(\rho _{0}^{z_{1}})d^{2}\rho
_{0}^{z_{1}}/dz_{1\,}^{2},  \label{mvdwrefode}
\end{equation}
where 
\begin{equation}
m_{\alpha }(\rho )\equiv m\alpha (\rho )/a.  \label{malpha}
\end{equation}

Following Rowlinson and Widom \cite{6}(RW) we can define 
\begin{eqnarray}
-W(\rho ) &\equiv &f(\rho )-\mu ^{\ell }\rho +p^{\ell },  \nonumber \\
&=&\rho [\mu (\rho )-\mu ^{\ell }]-[p(\rho )-p^{\ell }].  \label{Wdef}
\end{eqnarray}
Here $f(\rho )$ is the Helmholtz free energy density from our analytic
equation of state that corresponds to $\mu (\rho )$ given by Eq.~(\ref
{mualpha}), where $\mu (\rho )=df(\rho )/d\rho $ and the associated pressure 
$p(\rho )=-f(\rho )+\rho \mu (\rho )$ from standard thermodynamics. We see
that $W(\rho )$ vanishes in the coexisting bulk liquid and vapor phases and
note that the left side of Eq.~(\ref{mvdwrefode}) is given by $-dW(\rho
_{0}^{z_{1}})/d\rho _{0}^{z_{1}}.$ By interpreting the latter as a
``force'', Eq.~(\ref{mvdwrefode}) is analogous to Newton's law, with $\rho
_{0}^{z_{1}}$ the ``displacement'', $z_{1}$ the ``time'' and $m_{\alpha
}(\rho )$ a \emph{density} (or ``displacement'') \emph{dependent} ``mass''.

Note that Eq.~(\ref{mvdwrefode}) differs from the analogous equation that
would arise in the classical theory from assuming a density dependent
``mass'' in the gradient correction to the \emph{free energy}. As shown by
RW in their Eq.~(3.10), the latter would generate an additional square
gradient term in the interface equation (\ref{mvdwrefode}). It is difficult
to see how such a term could arise naturally in our approach.

We report results here for the even simpler theory that arises when $
m_{\alpha }(\rho )$ in Eq.~(\ref{mvdwrefode}) is replaced by its classical
value $m$ given by Eq.~(\ref{vdwm}). The resulting simplified interface
equation has the same form as the classical equation (\ref{vdwrefode}).
However it uses the accurate expression for $\mu (\rho )$ given by Eq.~(\ref
{mualpha}), which assures a proper thermodynamic description of the
coexisting bulk phases, while (inconsistently) retaining the classical
expression for $m.$ This produces some changes in shape of the interface
profile when compared to that predicted by the MVDW theory, but preserves
many qualitative features such as the dependence of the interface width on
the thermodynamic state and the range of the attractive interactions. One
major virtue of this approximation is that a very simple expression for the
profile $\rho _{0}^{z}$ in terms of the inverse function $z(\rho _{0})$
(with arbitrary origin) follows immediately from Eq.~(\ref{mvdwrefode})
by quadrature, as shown by RW: 
\begin{equation}
z(\rho _{0})-z(\rho ^{v})=\left( \frac{m}{2}\right) ^{1/2}\int_{\rho
^{v}}^{\rho _{0}}d\rho \,[-W(\rho )]^{-1/2}.  \label{zrho}
\end{equation}
Results from this simplified approach and the full MVDW theory will be
discussed in Sec. \ref{results} below.

\subsection{Nonlocal correlations and capillary waves}
The use of the HLR equation in the MVDW theory allows one to take account of
nonlocal correlations induced by the ERF. These arise mainly from the
packing of the harshly repulsive molecular cores and become significant at
high density when the ERF is rapidly varying. In principle small amplitude
excluded volume oscillations would be expected at high density from linear
response theory far from \emph{any} localized perturbation \cite
{evanslin1,evanslin2}. The amplitude of these oscillations for a general
liquid-vapor interface depends on the thermodynamic state and on the
strength and range of the attractive interactions. These features control
the steepness of the ERF $\phi _{R}(\mathbf{r),}$ which mainly
determines how significant these nonlocal corrections to the classical
theory are in a given case.

For our study here of the LV interface in the LJ fluid the attractive
interactions are relatively slowly varying and these corrections are
numerically small in most cases. However, near the triple point the
interface is very steep and noticeable oscillations on the liquid side are
predicted by the theory and, with much smaller amplitude, are also seen in
the computer simulations. The classical local hydrostatic approximation
precludes a description of any such oscillations and gives no indication of
where it can break down. So these corrections are conceptually important
even for ``smooth'' LV interfaces.

The difference in oscillation amplitude between theory and simulation arises
because the MVDW theory describes a ``static'' interface in the reference
fluid induced by the ERF. Any theory that takes account of attractive forces
only through an ERF and uses reference system correlation functions to
approximate structure in the LJ system cannot properly describe the physics
leading to the long-wavelength capillary wave fluctuations that occur at a
real LV interface \cite{6,BLS,15}. These induce characteristic long-ranged
pair correlations in the interface region of a real fluid that are
completely different from the corresponding pair correlations in the
reference system, which remain short-ranged for any reasonable choice of
$\phi _{R}(\mathbf{r})$. In a sufficiently large system, the capillary wave
fluctuations can wash out any excluded volume oscillations at the real LV
interface, and indeed the entire interface profile $\rho (\mathbf{r})$
itself \cite{BLS}!

However the small system sizes studied in computer simulations or
encountered in most experiments on interfaces in confined geometries cut off
most effects of such capillary wave fluctuations. It is reasonable to
interpret the MVDW theory as providing one way of defining an ``intrinsic''
profile unbroadened by capillary wave effects and to compare its predictions
directly to the simulation or experimental data, taking account of the
residual finite size capillary wave effects separately if necessary \cite{15}.
This will be discussed further below.

\begin{figure}[t]
\includegraphics[width=80mm,height=60mm]{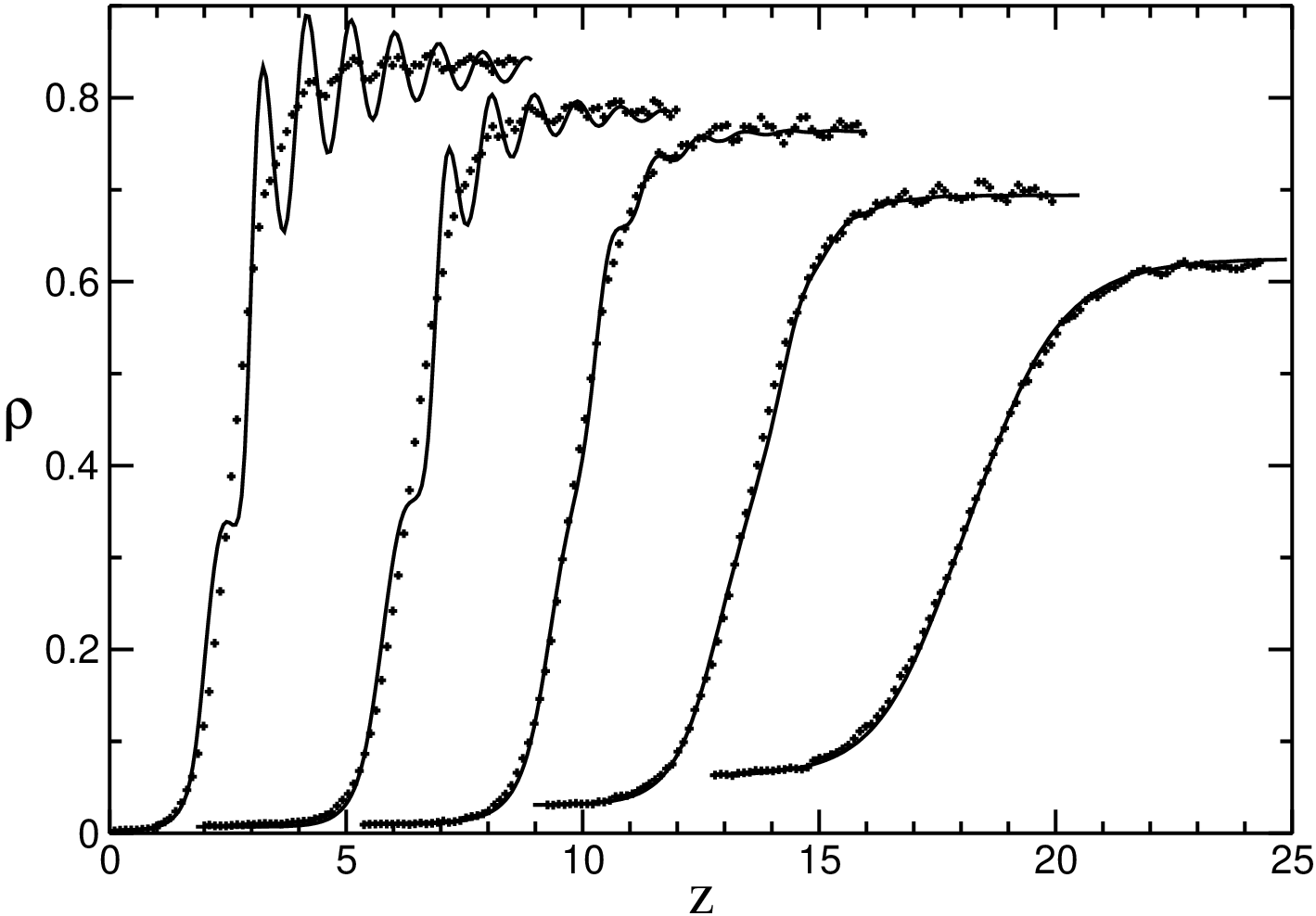}
\caption{\label{fig:mecke_SC} Density profiles of the LJ liquid-vapor interface
for states with different temperatures and potential energy cutoffs.
From left to right: $T=0.7$ and $r_c=5.0$; $T=0.7$ and $r_c=2.5$;
$T=0.85$ and $r_c=5.0$; $T=0.85$ and $r_c=2.5$;
$T=1.1$ and $r_c=5.0$. Symbols are results of Mecke et al. \cite{7}
Lines are predictions of the MVWD theory.}
\end{figure}

Workers using other approaches such as density functional theory or integral
equation methods sometimes argue that their theories describe correlations
in the full nonuniform LJ system, and thus may include some or perhaps all
of the effects of capillary wave fluctuations \cite{16,17}. However such
theories usually introduce approximations that relate correlation functions
in the nonuniform system to interpolated or weighted correlation functions
in the uniform LJ system. One must then deal with the ambiguities arising from
unstable uniform densities in the two phase region. We believe most such
arbitrary schemes implicitly introduce a mean field character to the theory
through the use of uniform fluid correlation functions that do not contain
any capillary wave effects. However the precise physical implications of
such approximations are very difficult to assess. In our interpretation of
MF theory, both the strengths and the limitations arising from the use of
reference system correlation functions are clear from the outset.

\section{Results for interface structure}
\label{results}

\subsection{MVDW theory}

Figure~1 shows the interface profiles for the LJ fluid for states at three
different temperatures. The dots give results of recent molecular dynamics
(MD) simulations of Mecke et al. \cite{7}. They made a careful study of the
important changes in the interface profile and surface tension that arise
from setting the force from the LJ potential to zero beyond a certain cutoff
distance $r_{c}$. Since there are unbalanced attractive forces in the
interface region, effects from different cutoffs
are much more important than in uniform
systems, where the attractive forces essentially cancel. The first two
curves on the left give profiles very near the triple point at a reduced
temperature $T$ of $0.7$ with $r_{c}=5$ and $2.5$ respectively. There are
notable changes in the coexistence densities $\rho ^{v}$ and $\rho ^{\ell }$
from the different cutoffs. The next two curves give results at $T=0.85$
with the same two cutoffs and the last curve on the right gives $T=1.1$ with 
$r_{c}=5.$

The lines give results of the MVDW theory, using the appropriate cutoff and
shifted LJ potential. The effects of the cutoff on the bulk thermodynamics
can be taken into account in the general equation of state of Johnson, et
al. \cite{13} used in the MVDW theory, and the ERF properly describes the
averaged effects of the unbalanced attractive forces in the interfacial
region. Overall there is very good agreement between simulations and the
theory, which captures all qualitative effects of changes in temperature and
cutoff radius.

However, the theory predicts very noticeable density oscillations on the
liquid side at the lowest temperature $T=0.7$. As discussed earlier, the
amplitude of these oscillations is a sensitive function of temperature and
cutoff and already by $T=0.85$ their amplitude is greatly reduced. Such
nonlocal excluded volume correlations are to be expected when the density is
high and the ERF is sufficiently rapidly varying. In other contexts they
play an important role in the physics of nonuniform fluids \cite{4}.

Indeed we find that the theory provide an exceptionally accurate description
of $\rho _{0}(\mathbf{r};[\phi _{R}],\mu _{0}^{\ell }),$ the density of the 
\emph{reference system} in the presence of the ERF $\phi _{R}$. This is
illustrated in Fig.~2, which compares the theory (solid lines)\ for $T=0.7$
and $T=0.75$ directly to results from grand canonical Monte Carlo
simulations we carried
out \cite{sim} for the \emph{reference system} density (symbols) in the
presence of the self-consistently determined ERF (dotted lines). Note that
while the field is smooth, it is sufficiently rapidly varying in this case
to produce a density response with nonlocal excluded volume correlations.
These are very well described theoretically by the HLR equation used in the
MVDW theory and are completely missed by the local hydrostatic approximation
(dashed lines) used in the classical theory.

The differences in oscillation amplitude seen in Fig.~1 arise because the
reference system cannot describe the capillary wave fluctuations seen at a
real LV interface, as discussed above. Some residual effects are present
even for the relatively small system sizes used in the computer simulations.
We can take them into account in an approximate way by using the standard
prescription for Gaussian capillary wave smearing of an ``intrinsic''
interface. Thus we convolute the ``intrinsic''
interface as given by the MVDW theory in Fig.~1 with a Gaussian
distribution of local interface positions $h$: 
\begin{equation}
P(h)=(2\pi s^{2})^{-1/2}\exp (-h^{2}/2s^{2}),  \label{gaussianp}
\end{equation}
where the width $s$ of the Gaussian is given by
\begin{equation}
s^{2}=\frac{k_{B}T}{2\pi \gamma }\ln \frac{L}{L_{s}}.  \label{gaussainwidth}
\end{equation}
Here $L$ is the lateral box size in the simulation and the $L_{s}$ is the
short distance (wavelength) cutoff, which is supposed to be proportional to
the bulk correlation length or the interface width \cite{6,BLS,15}.
In this case the amount of smoothing depends mainly on the choice of $L_{s}$,
and the very reasonable
choice of three times the interface width gives the data plotted in Fig.~3.
The minimal choice of just the interface width gives too much smoothing
when compared to the MD data.

\begin{figure}[t]
\includegraphics[width=80mm,height=60mm]{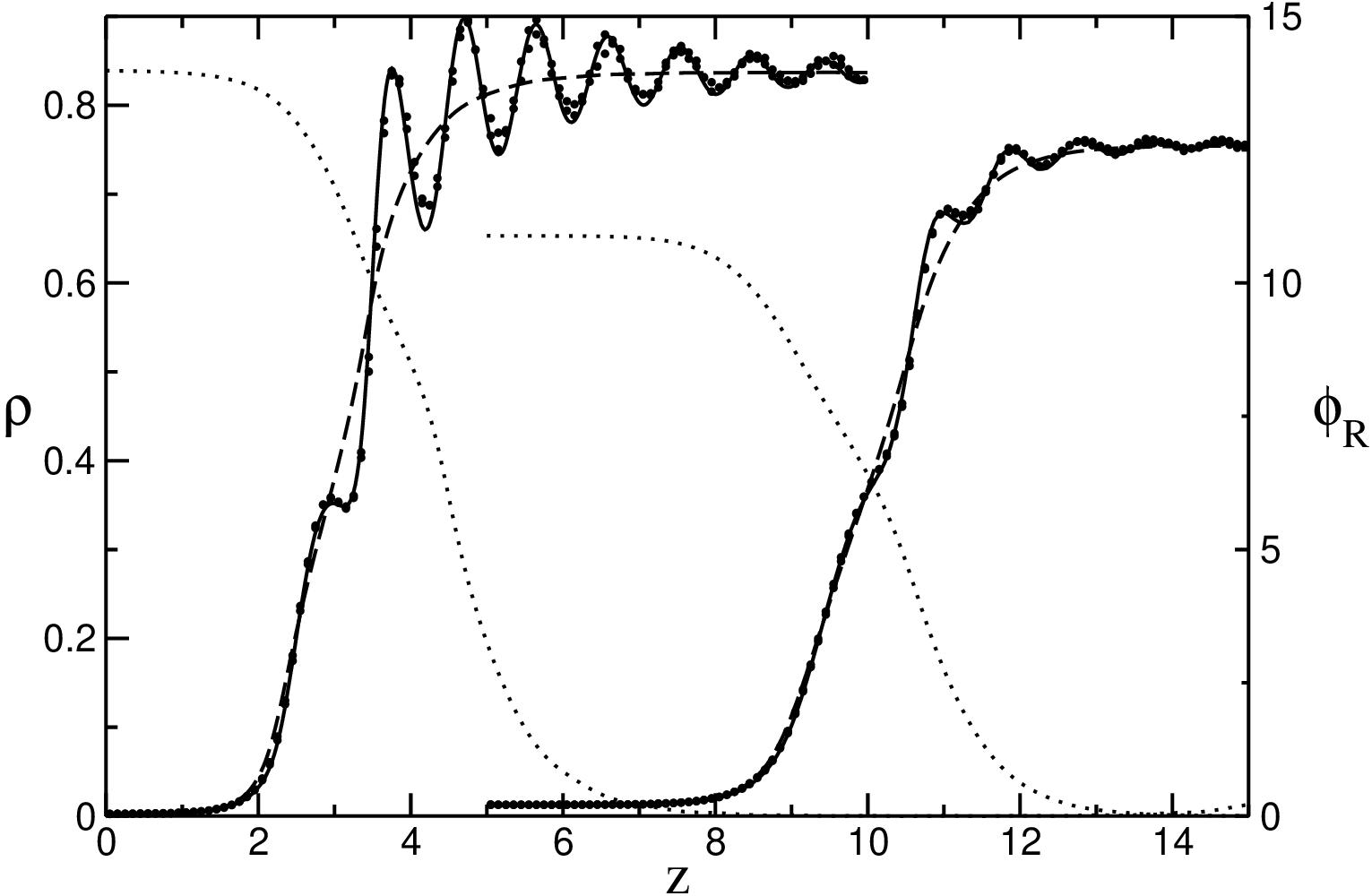}
\caption{\label{GCMC_ref} MC simulation of the \emph{reference system}
in the presence of the
self-consistent field $\phi _R$, given by the dotted lines (use the right vertical
axis for fields). The full lines and symbols denote
densities as in Fig.~1. The dashed lines gives the hydrostatic density defined
by Eq.~(\ref{hydromudef}). The states are: (left) $T=0.7$, $r_c=5.0$;
(right) $T=0.75$, $r_c=2.5.$}
\end{figure}

We see that the finite size capillary wave fluctuations have very little
effect on the smooth profiles at higher temperatures, but are quite
effective in damping out some of the ``intrinsic'' oscillations predicted
near the triple point. Our purpose here is not to advocate this particular
and somewhat arbitrary prescription for smoothing the results of the MVDW
theory, but to point out that while excluded volume oscillations are
accurately described by the MVDW theory, their influence on the LV interface
profile is reduced by capillary wave effects not captured by the theory. In
many other applications, e.g., structure near a hard wall or in a slit or
pore, the capillary wave fluctuations are suppressed even further and the
MVDW theory gives a good description of the excluded volume correlations
such rapidly varying fields induce \cite{4}.

\subsection{Simplified MVDW theory}
Figure~4 gives the interface profiles predicted by the simplified MVDW
theory from Eq.~(\ref{zrho}). The theory seems to give very good results,
perhaps even better than those of the full MVDW theory in Fig.~1! However
part of this agreement is a result of a fortuitous cancellation of errors.
As shown in Fig.~2, the
hydrostatic approximation used in the simplified theory completely
suppresses the excluded volume correlations that should really be present at 
$T=0.7$ in the reference system profile. As a result it seems to give better
agreement with the simulation data for the LJ profile, where capillary wave
effects also not properly described by the simplified theory wash out most
effects of the oscillations. Note that the simplified theory gives least
accurate results at $T=1.1$ where the interface is more slowly varying, and
one would have expected the theory to be most accurate. As we will see later,
results for the surface tension from the simplified theory are also less
satisfactory. Nevertheless the simplified theory captures well the
qualitative effects of temperature and cutoff on the interface profile and
it is exceptionally easy to implement.

\begin{figure}[t]
\includegraphics[width=80mm,height=60mm]{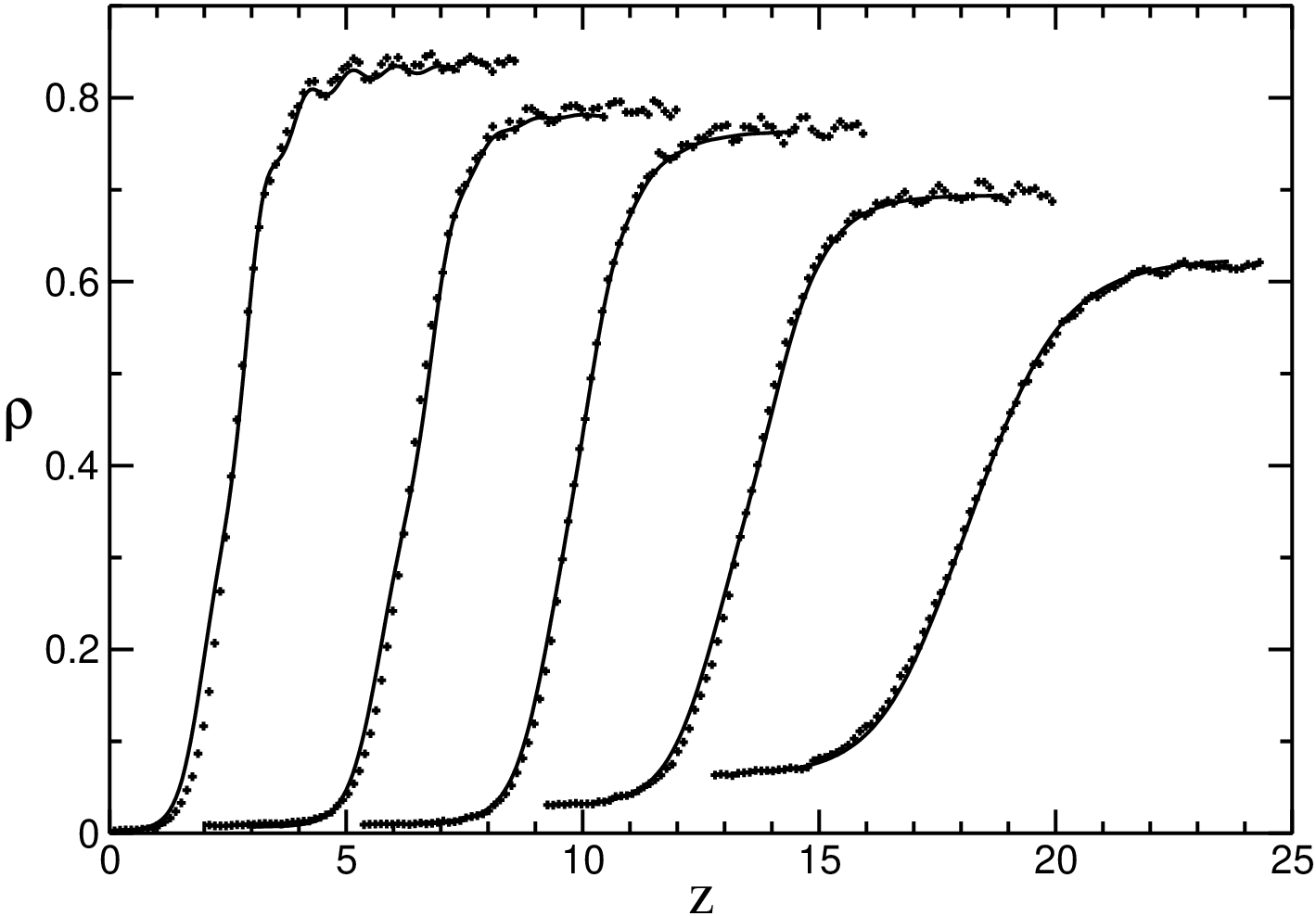}
\caption{\label{mecke_SC_cw3} MVDW theory with capillary wave smoothing
of profiles in Fig.~1, with $L_s$
equal to three times the interface width.}
\end{figure}

\section{Surface tension}
\subsection{Basic formalism}
In the MVDW theory, we first determine fluid structure. To calculate the
interface free energy or surface tension $\gamma $ for the LV system, we
proceed formally and imagine a special path where the density in the LJ
fluid changes linearly \cite{otherpaths} from that of the uniform liquid
with density $\rho ^{\ell }$ to the final LV interface profile $\rho (%
\mathbf{r};\mathbf{[}\phi =0],\mu ^{\ell })\equiv \rho (\mathbf{r)}$ as
controlled by a coupling parameter $\lambda $ with $0\leq \lambda \leq 1$: 
\begin{equation}
\rho _{\lambda }(\mathbf{r})=\rho ^{\ell }+\lambda [\rho (\mathbf{r})-\rho
^{\ell }].  \label{rholambda}
\end{equation}
Here $\rho _{\lambda }(\mathbf{r})\equiv \rho (\mathbf{r;[}\phi _{\lambda
}],\mu ^{\ell })$ where $\phi _{\lambda }(\mathbf{r})$ is the (generally
nonzero)\ external field that formally produces the partially coupled
profile $\rho _{\lambda }(\mathbf{r})$ defined by the right side of
Eq.~(\ref{rholambda}). Since
$\rho _{\lambda }({\bf r})=\delta \Omega _{\lambda }/\delta
\phi _{\lambda }({\bf r})$,
on integration the change in the Grand canonical free energy associated with
this density change is exactly given by: 
\begin{equation}
\Omega _{\lambda =1}-\Omega _{\lambda =0}=\int d\mathbf{r}
\int_{0}^{1}d\lambda \,\rho _{\lambda }(\mathbf{r})\frac{d\phi _{\lambda }(%
\mathbf{r})}{d\lambda }.  \label{delomega}
\end{equation}
Here $-\beta \Omega _{\lambda }\equiv \ln \Xi _{\lambda }$ where $\Xi
_{\lambda }$ is the Grand partition function for the system with field $\phi
_{\lambda }.$ For the LV interface we have $\phi _{\lambda =0}(\mathbf{r}
)=\phi _{\lambda =1}(\mathbf{r})=0.$ Since $\Omega _{\lambda =0}=-p^{\ell }V$,
with $p^{\ell }$ the coexistence pressure, which equals that in the vapor
phase, the free energy difference on the left side is the desired
interfacial free energy $\gamma ,$ and is independent of any choice of Gibbs
dividing surface \cite{6}. Integrating by parts, we have our basic starting
point: 
\begin{equation}
\gamma =-\int d\mathbf{r}\int_{0}^{1}d\lambda \,\phi _{\lambda }(\mathbf{r})%
\frac{d\rho _{\lambda }(\mathbf{r})}{d\lambda },  \label{gamma}
\end{equation}
or for the linear path: 
\begin{equation}
\gamma =-\int d\mathbf{r}[\rho (\mathbf{r})-\rho ^{\ell
}]\int_{0}^{1}d\lambda \phi _{\lambda }(\mathbf{r}).  \label{gammalin}
\end{equation}

\begin{figure}[t]
\includegraphics[width=80mm,height=60mm]{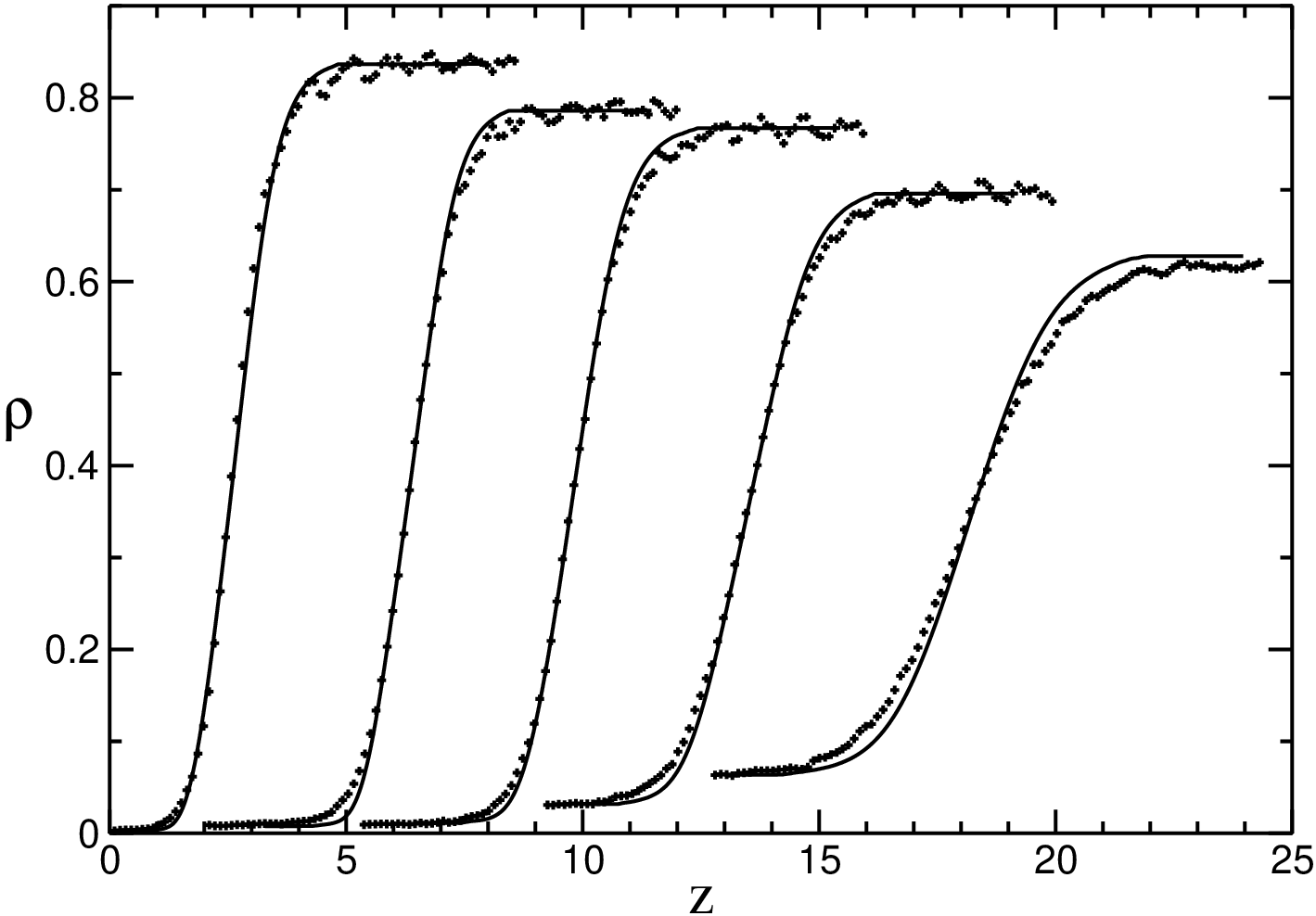}
\caption{\label{mecke_VDW} Simplified MVDW density profiles from Eq.~(16).
Same states as in Fig.~1}
\end{figure}

\subsection{MVDW theory for surface tension}
The MF approximation used in the MVDW theory allows us to evaluate
these expressions using reference system quantities. Thus we assume from
Eq.~(\ref{singletden}) that $\rho _{\lambda }(\mathbf{r})=\rho _{0\lambda }(%
\mathbf{r} )\equiv \rho _{0}(\mathbf{r;[}\phi _{R\lambda }],\mu _{0}^{\ell
}),$ where $\phi _{R\lambda }(\mathbf{r})$ is the field in the reference
system producing the same profile, formally related to $\phi _{\lambda }(%
\mathbf{r)}$ by the MMF equation (\ref{mmfint}): 
\begin{eqnarray}
\phi _{R\lambda }({\bf r}_{1}) &=&\phi _{\lambda }({\bf r}_{1})+\frac{\alpha
(\rho _{0\lambda }^{{\bf r}_{1}})}{a}\int d{\bf r}_{2}\,\rho _{0\lambda }
({\bf r}_{2})\,u_{1}(r_{12})  \nonumber \\
&&+2\alpha (\rho ^{\ell })\rho ^{\ell }.  \label{mmintlambda}
\end{eqnarray}
Using Eq.~(\ref{hydromudef}) and Eq.~(\ref{mualpha}), we can
exactly rewrite Eq.~(\ref{mmintlambda}) in a convenient form
for use in Eqs.~(\ref{gamma}) and (\ref{gammalin}): 
\begin{equation}
\phi _{\lambda }(\mathbf{r}_{1})=\mu ^{\ell }-\mu (\rho _{0\lambda }^{%
\mathbf{r}_{1}})-\frac{\alpha (\rho _{0\lambda }^{\mathbf{r}_{1}})}{a}\int d%
\mathbf{r}_{2}\,[\rho _{0\lambda }(\mathbf{r}_{2})-\rho _{0\lambda }^{%
\mathbf{r}_{1}}]\,u_{1}(r_{12}).  \label{philambda}
\end{equation}
In the MVDW theory $\mu (\rho )$ is determined from the accurate equation of
state and is given by Eq.~(\ref{mualpha}), and all densities are calculated
in the reference system.

To calculate $\phi _{\lambda }(\mathbf{r}_{1})$ from Eq.~(\ref{philambda})
we start with the final self-consistent profile $\rho _{0}(\mathbf{r)}$ and
use Eq.~(\ref{rholambda}) to define $\rho _{0\lambda }(\mathbf{r}).$ We then
iterate the HLR equation (\ref{HLR}) in an ``inverse'' way to find the
hydrostatic density $\rho _{0\lambda }^{\mathbf{r}_{1}}$ associated with a
given $\rho _{0\lambda }(\mathbf{r}_{1})$ (Although we know that $\rho
_{0\lambda }(\mathbf{r)}$ is linear in $\lambda $ from Eq.~(\ref{rholambda}%
), unless the density is slowly varying this does not imply that the same
condition holds for $\rho _{0\lambda }^{\mathbf{r}}.$) With this in hand, we
use Eq.~(\ref{philambda}) to determine $\phi _{\lambda }(\mathbf{r}_{1})$
for several intermediate values of $\lambda .$ The surface tension is then
calculated by carrying out the integration in Eq.~(\ref{gammalin})
numerically.

\subsection{Simplified hydrostatic approximations for the surface tension}
If we ignore the difference between $\rho _{0\lambda }(\mathbf{r}_{1})$ and $
\rho _{0\lambda }^{\mathbf{r}_{1}}$ as in the classical theory and assume
the latter varies linearly in $\lambda ,$ then we require only the final
profile $\rho _{0}^{\mathbf{r}_{1}}$ and can carry out most of the $\lambda $
integration in Eq.~(\ref{gamma}) analytically. We have already seen that this is a
rather accurate approximation for the structure of the LV interface and we
use this approximation only under an integral in computing the free energy.
As we will see, this greatly simplifies the calculation of the free energy
and also allows us to make contact with classical results and the simplified
MVDW theory.

Assuming that $\rho _{0\lambda }(\mathbf{r}_{2})=\rho _{0\lambda }^{\mathbf{r%
}_{2}} $ in Eq.~(\ref{philambda}) we have 
\begin{equation}
\phi _{\lambda }(\mathbf{r}_{1})=\mu ^{\ell }-\mu (\rho _{0\lambda }^{%
\mathbf{r}_{1}})-\frac{\alpha (\rho _{0\lambda }^{\mathbf{r}_{1}})}{a}\int d%
\mathbf{r}_{2}\,[\rho _{0\lambda }^{\mathbf{r}_{2}}-\rho _{0\lambda }^{%
\mathbf{r}_{1}}]\,u_{1}(r_{12}),  \label{philambdainthydro}
\end{equation}
and we will use this expression in Eqs.~(\ref{gamma}) or (\ref{gammalin}) to
determine the surface tension. For $\lambda =1$ we have $\phi _{\lambda =1}(%
\mathbf{r}_{1})=0,$ and Eq.~(\ref{philambdainthydro}) reduces to the
generalized interface equation (\ref{mvdwrefode}) on expanding $\rho
_{0\lambda }^{\mathbf{r}_{2}}$ to second order about $\rho _{0\lambda }^{%
\mathbf{r}_{1}}.$ The simplified MVDW interface equation discussed earlier
whose solution is given in Eq.~(\ref{zrho}) follows on further approximating 
$\alpha (\rho )$ by $a$ in Eq.~(\ref{philambdainthydro}) or $m_{\alpha }$
by $m$ in Eq.~(\ref{mvdwrefode}).

Now let carry out the $\lambda $ integration in Eq.~(\ref{gamma}) using Eq.~(%
\ref{philambdainthydro}) for $\phi _{\lambda }(\mathbf{r}_{1}).$ A clear
discussion of most of the technical issues in given by RW. The first two
terms on the right in Eq.~(\ref{philambdainthydro}) represent the local
contribution to the interface free energy and can be integrated
analytically: 
\begin{eqnarray}
I_{local} &\equiv &-\int d\mathbf{r}\int_{0}^{1}d\lambda \frac{d\rho
_{0\lambda }^{\mathbf{r}}}{d\lambda }\mathbf{[}\mu ^{\ell }-\mu (\rho
_{0\lambda }^{\mathbf{r}})]  \nonumber \\
&=&-\int d\mathbf{r}\int_{\rho ^{\ell }}^{\rho _{0}^{\mathbf{r}}}d\rho
_{0\lambda }^{\mathbf{r}}\frac{dW(\rho _{0\lambda }^{\mathbf{r}})}{d\rho
_{0\lambda }^{\mathbf{r}}}  \nonumber \\
&=&-\int d\mathbf{r}W(\rho _{0}^{\mathbf{r}}),  \label{Ilocal}
\end{eqnarray}
where $W(\rho )$ is given in Eq.~(\ref{Wdef}).

\begin{figure}[t]
\includegraphics[width=80mm,height=60mm]{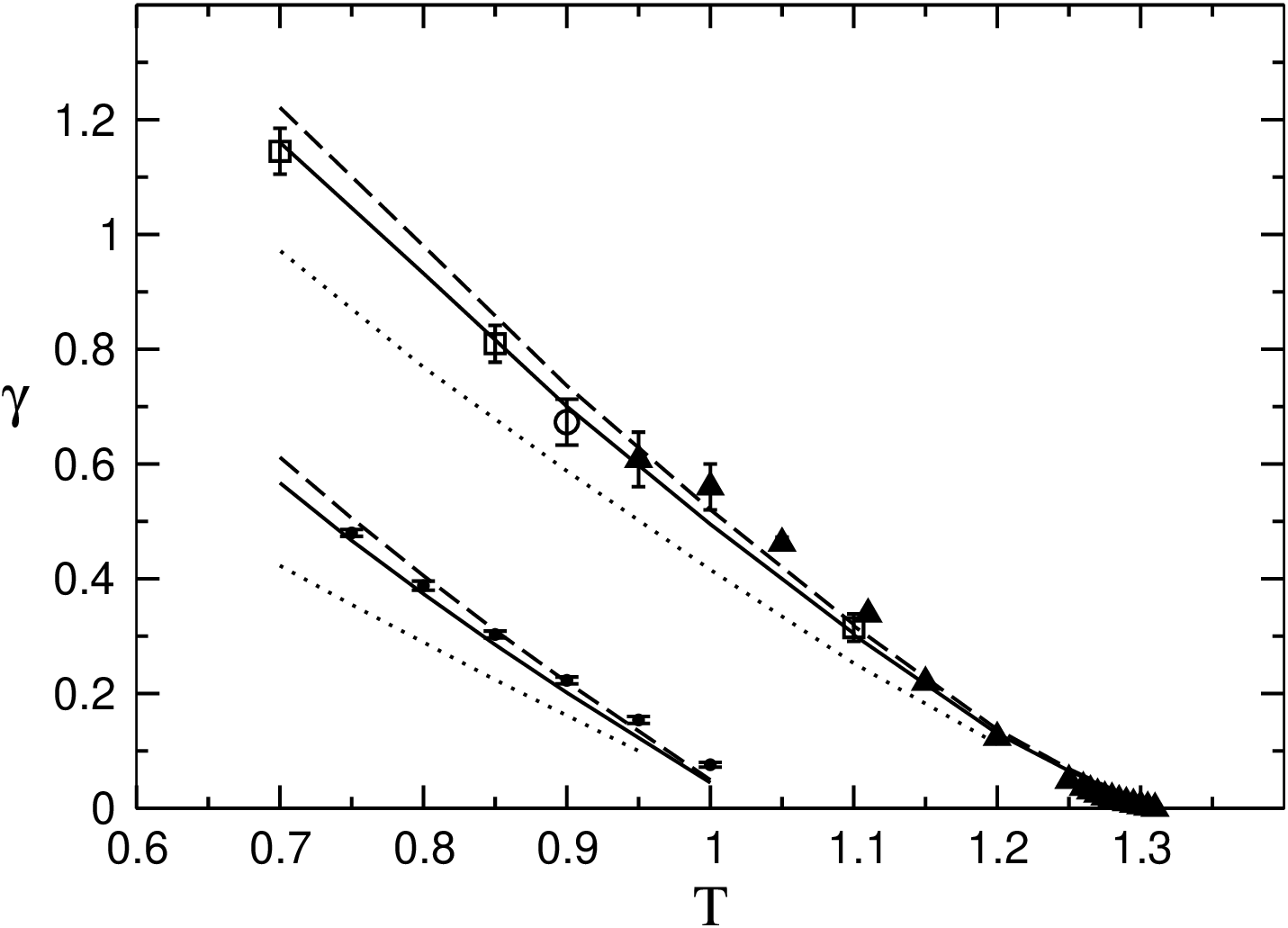}
\caption{\label{st_all} Surface tension $\gamma$ of
the LJ liquid-vapor interface.
The two sets correspond to the full LJ interaction potential (upper) and
the cut-and-shifted LJ interaction with $r_c=2.5$ (lower).
Full lines are MWVD predictions. Dashed lines use the hydrostatic
density to compute $I_{nonlocal}$ from Eq.~(\ref{Inonl}).
Dotted lines are results of the simplified MVDW theory from Eq.~(\ref{gammaW}).
Results of simulations are from Mecke et al. \cite{7} (squares),
Holcomb et al. \cite{7a}
(empty circle), Haye et al. \cite{7b} (filled circles) and Potoff et al.
\cite{7c} (filled triangles).}
\end{figure}

The last term in Eq.~(\ref{philambdainthydro}) gives the nonlocal
contribution to the free energy. Using Eq.~(\ref{rholambda}) for $\rho
_{0\lambda }^{\mathbf{r}}$ we require only $\rho _{0}^{\mathbf{r}}$ and from
Eq.~(\ref{gammalin}) we find 
\begin{equation}
I_{nonlocal}=\int d\mathbf{r}_{1}\mathbf{\,}[\rho _{0}^{\mathbf{r}_{1}}-\rho
^{\ell }]K(\mathbf{r}_{1}\mathbf{)}\int d\mathbf{r}_{2}\,[\rho _{0}^{\mathbf{
r}_{2}}-\rho _{0}^{\mathbf{r}_{1}}]\,u_{1}(r_{12})\,  \label{Inonl}
\end{equation}
where 
\begin{equation}
K(\mathbf{r)}\equiv 2\int_{0}^{1}d\lambda \,\lambda \alpha (\rho _{0\lambda
}^{\mathbf{r}})/a.  \label{Kdef}
\end{equation}
Thus we have 
\begin{equation}
\gamma =I_{local}+I_{nonlocal}\,,  \label{gammaI}
\end{equation}
and both terms require only $\rho _{0}^{\mathbf{r}}.$ In the simplified MVDW
theory discussed earlier, we set $\alpha =a$ or $K(\mathbf{r})\mathbf{=}1$
in Eq.~(\ref{Inonl}) and expand $\rho _{0}^{\mathbf{r}_{2}}$ about $\rho
_{0}^{\mathbf{r} _{1}}$ to second order. Using Eq.~(\ref{mvdwrefode}) with $
m_{\alpha }=m$, we see that these additional approximations imply that $
I_{nonlocal}=$ $I_{local}$, as given in Eq.~(\ref{Ilocal}).
RW show in this case we can use the even simpler expression 
\begin{equation}
\gamma =\int_{\rho ^{v}}^{\rho ^{\ell }}d\rho [-2mW(\rho )]^{1/2},
\label{gammaW}
\end{equation}
which does not require explicit knowledge of $\rho _{0}^{\mathbf{r}}.$

\subsection{Results}
Figure 5 gives the surface tension predicted by the MVDW theory (with no
capillary wave smoothing of the profile), that arising from use of the
hydrostatic approximation as in Eq.~(\ref{Ilocal}) and (\ref{Inonl}), and
that given by the simplified MVDW theory in Eq.~(\ref{gammaW}). We see that
the MVDW theory gives very good agreement with the simulation data. The
simplified MVDW theory is much less accurate. Since all theories give
essentially the
same results for $I_{local}$, the problem with the simplified theory must
arise from its treatment of nonlocal effects through the approximation
$I_{nonlocal}=I_{local},$ which we see becomes increasingly inaccurate at
lower temperatures. Equation (\ref{Inonl}) provides a more accurate but
still simple alternative to use of the full MVDW theory.

\section{Conclusions}
The MVDW theory provides a simple and physically motivated approach that
can describe the structure and thermodynamics of a fluid in a general
external field. It optimally combines two standard approximations, a
molecular field treatment of attractive interactions, modified to give
accurate thermodynamic data for the uniform fluid, along with a linear
response treatment of correlations in the reference fluid. The present
application of the theory to the LV interface permits a new interpretation
of the classical VDW theory that removes some ambiguities in standard
treatments and shows how key features of the classical theory can be
improved in a natural way. The accuracy of the MVDW theory in this
application provided additional support for the physical ideas behind the
theory and for its quantitative utility.

It is a pleasure to dedicate this paper to John Tully on the happy occasion
of his 60th birthday. This work is supported by the National Science
Foundation through Grant CHE-0111104. We are grateful to Yng-Gwei Chen and
Jim Henderson for helpful discussions, and thank J. Winkelmann for sending
us his computer simulation data.


\begin{thebibliography}{99}
\bibitem{sim}  We carried out a standard Grand Canonical MC simulation.
There are three types
of moves: insertion of a particle, removal of a particle and a Canonical
local movement of a particle. The ratio of the moves was 1:1:20, which we did
not try to optimize.
Both a hard sphere fluid of appropriate diameter in the field and
and the reference LJ fluid in the same field were studied. Both
simulations give
essentially indistinguishable results. The hard sphere simulation is
considerably faster and the figure has those results.
\end{thebibliography}
\end{document}